\documentstyle[prl,aps,preprint,tighten,floats,aps,epsf,psfig]{revtex}

\newif\iftightenlines\tightenlinesfalse
\tightenlines\tightenlinestrue

\begin{document}
%
\def\VEV#1{{\langle #1 \rangle} }
\def\ord#1{{\cal{O}}(#1) }
\def\gsim{\lower.7ex\hbox{$\;\stackrel{\textstyle>}{\sim}\;$}}
\def\lsim{\lower.7ex\hbox{$\;\stackrel{\textstyle<}{\sim}\;$}}
\def\mol{Mol}
\def\etmiss{E\llap/_T}
\def\eslt{E\llap/_T}
\def\esl{E\llap/}
\def\msl{m\llap/}
\def\to{\rightarrow}
\def\te{\tilde e}
\def\tmu{\tilde\mu}
\def\ttau{\tilde\tau}
\def\tl{\tilde\ell}
\def\ttau{\tilde \tau}
\def\tg{\tilde g}
\def\tnu{\tilde\nu}
\def\tell{\tilde\ell}
\def\tq{\tilde q}
\def\tu{\tilde u}
\def\tc{\tilde c}
\def\ts{\tilde s}
\def\tb{\tilde b}
\def\tst{\tilde t}
\def\tt{\tilde t}
\def\tw{\widetilde W}
\def\tz{\widetilde Z}

\hyphenation{mssm}
%
\preprint{\vbox{\baselineskip=14pt%
   \rightline{UM-TH-99-09}\break 
   \rightline{hep-ph/9911243}\break 
}}
\title{Effects of Large CP-violating Soft Phases on Supersymmetric Electroweak
Baryogenesis}
\author{Michal Brhlik, Gerald J. Good and G.L. Kane}
\address{
Randall Laboratory of Physics,
University of Michigan,
Ann Arbor, MI 48109 USA
}
\date{\today}
\maketitle
\begin{abstract}

We revisit the results of recent electroweak baryogenesis calculations
and include all allowed large CP-violating supersymmetric phases. 
If the phases are large, the resulting baryon asymmetry can be considerably
larger than the observed value $n_B/s \sim 4 \times 10^{-11}$.  
Much of the asymmetry must therefore be washed out, and we argue that
the upper bound on the light Higgs mass is larger than the value
reported in previous work.
   
\end{abstract}

\medskip



\section{Introduction}

The universe is not matter-antimatter symmetric; as shown by primordial 
nucleosynthesis measurements, the baryon density to 
entropy density ratio $n_B/s$ is constrained to be about
$ 4 \times 10^{-11}$ \cite{steig}.  The necessary ingredients for a theory 
to explain this asymmetry are the Sakharov criteria: baryon number violation, 
C and CP violation and non-equilibrium conditions \cite{sakh}.
Electroweak baryogenesis is an attractive mechanism with the best chance 
of verification through traditional collider experiments \cite{review}.

The central feature of electroweak baryogenesis is spontaneous symmetry
breaking.  At high temperatures, the vaccum expectation value of the 
Higgs field is zero but as the universe cools down over time, a second
minimum appears in the potential at $v \neq 0$.  The universe
undergoes a phase transition, from a ``symmetric'' phase where $v = 0$ to
the ``broken'' phase where $v \neq 0$.  If the phase transition is second
order, which means that no potential barrier exists, then the universe
is approximately in equilibrium throughout the transition and no 
baryogenesis can result (since the third Sakharov criterion is violated).
If the phase transition is first order, then the phase transition occurs
through quantum tunneling. As the universe cools, the probability of
making a transition from symmetric to broken phase grows and when the
transition occurs at a certain point in space, a bubble of broken phase 
forms and expands.  Because of its first order nature, the phase 
transition does not occur throughout the whole universe at the same
time,  resulting in non-equilibrium conditions.

The baryon asymmetry is generated as the wall of the expanding bubble
passes through points in space. Particles in the unbroken phase close 
to the wall interact with the changing Higgs field profile in the
wall and the presence of CP-violating couplings produces source terms
for participating particles. Different chiralities couple with different
strength when CP is violated and a difference occurs in the
reflection and transmission probabilities for the two different chiralities.
Due to rapid gauge, Yukawa and 
strong sphaleron interactions, the CP-violating source terms are 
translated into a net left handed weak doublet quark density which is
finally converted into a baryon asymmetry by weak sphaleron decays. The
asymmetry then diffuses through the bubble wall into the broken phase
where the weak sphaleron interactions are exponentially suppressed.
Subsequent washout of the baryon asymmetry can therefore be kept under 
control provided the first order phase transition is strong enough.
    
The Standard Model satisfies the three Sakharov criteria, but it can not
generate the required value of $n_B/s$.  This is because the only
source of CP violation in the SM comes from the phase $\delta$ in the 
CKM quark mixing matrix.  The baryon asymmetry is $\sim \sin \delta$, but
this value is suppressed by flavor mixing factors \cite{ghopq,hs} and
the resulting asymmetry cannot be too large.  Therefore, the phase transition
has to be strongly first order to avoid washing out the produced baryon
number, which translates into the criterion
$v(T_c)/T_c \gsim 1$ \cite{krs}.  However, in the Standard Model, the
phase transition is too weakly first order to prevent washout,
unless the SM Higgs mass is less than 50 GeV, far below the present
experimental limit \cite{fkrs}.

The situation can be improved in the Minimal Supersymmetric Standard
Model (MSSM) as the light right handed top squark contribution to the
temperature dependent effective potential can push the value of
$v(T_c)/T_c$ up to acceptable values even for light Higgs masses allowed  
by experimental searches
\cite{beqz,cqw1,dgg-fw,espy1,laine,clinekain,moroakq,losada,bjls,espydc,lainerum,climoo}.  
Also, supersymmetric extensions of the SM include additional sources, 
which in general can
further enhance the possibilty of baryon asymmetry production during the
electroweak phase transition.  As a result, a specific region in the MSSM
parameter space corresponding to a heavy CP-odd Higgs boson, a light
CP-even  Higgs boson, a light right-handed stop and a heavy left-handed
stop can provide a plausible framework for electroweak baryogenesis.

The CP-violating interactions in the MSSM arise in the complex phases of
the soft supersymmetry breaking terms in the Lagrangian and in the phase
of $\mu$. Since the  largest contributions come from charginos and 
neutralinos \cite{huetnel,cqrvw} and less importantly from the right
handed stops, only the gaugino mass phases $\varphi_1$, $\varphi_2$ and 
$\varphi_3$, the phase of $\mu$ and the
phase of the stop trilinear parameter $A_t$ are relevant for
baryogenesis. These phases have to be small, typically $\lsim 10^{-2}$, if
they are considered individually with sparticle masses $\cal O$(TeV),
otherwise they induce contributions to the electric dipole moments
(EDMs) of the neutron and electron exceeding experimental limits 
\cite{oldedm,edm,gar}. However, these constraints can be avoided if 
relations among soft breaking parameters ensure cancellations of
individual contributions to the EDMs \cite{nath,bgk}. In the most
general case this possibility leads to models with light
superpartner spectra and CP-violating phases of $\cal O$(1) \cite{bgk}.
Remarkably, string motivated scenarios with non-universal gaugino masses 
can be described where such EDM cancellations occur naturally as a result
of the SM embedding on five-branes \cite{bekl}.      
   
Different methods of calculating the
baryon asymmetry, using both classical and quantum Boltzmann equations, have
produced the required asymmetry, provided that $\sin \varphi_{\mu} \sim
10^{-2} - 10^{-4}$ \cite{huetnel,cqrvw,aos,clijoyce,riotto,riussanz} 
in agreement with the experimental limits on the EDMs of the electron
and  neutron, when the other MSSM phases are taken to be zero. 
In this paper we assume that large phases are allowed by the cancellation
mechanism and at the same time the right handed stop is
light enough to significantly modify the finite temperature Higgs
potential. All other sfermions are heavier and their contribution to the Higgs 
potential is Boltzmann suppressed. 

We briefly review the generation of the baryon 
asymmetry and the phase  dependence factorization emphasizing the
significance of the $\varphi_2+\varphi_{\mu}$ phase combination.
In the large phase scenario it is then possible that the amount of
baryon asymmetry 
generated at the phase transition is $10^2 - 10^4$ larger than what is
observed.

We solve the full quantum Boltzmann equations to extract the CP
violating  source terms which also provides enhancement compared to
classical Boltzmann equations due to quantum memory effects \cite{riotto}.  

It is necessary to wash out some of the asymmetry;
therefore the constraint $v(T_c)/T_c \gsim 1$ can be relaxed.
We reevaluate the washout calculation using the effects
of requiring some washout to occur and derive our limits for the light
Higgs mass and the right handed stop based on electroweak baryogenesis.  
We are able to  show that even in the one-loop approximation for the 
temperature dependent effective Higgs potential the light Higgs mass 
upper limit can be pushed up to 115 GeV without invoking negative values
of the right handed stop soft breaking mass parameter $m_{U}^2$, which can
lead to color-breaking global minima and scalar potential
instability \cite{cqw1}.

\section{Baryon Asymmetry Calculation}  
 
In order to calculate the baryon asymmetry of the universe, one has to
start with a self-consistent computation of the CP-violating sources
resulting from particle interactions with the changing Higgs profile in
the bubble wall. We follow the procedure of reference \cite{riotto},
which uses the closed-time-path formalism of finite-temperature field
theory to derive quantum Boltzmann equations for Higgsinos and stops.
Quantum memory effects resulting from correct treatment of particle
propagation in the plasma strengthen the non-equilibrium character of
the the particle scattering off the bubble wall and produce larger
source terms. 

The dominant sources of CP-violation that are relevant to the electroweak
baryogenesis scenario comes from the interactions of the Higgs fields 
with charginos and neutralinos. These interactions couple the Higgsino
and gaugino components of charginos and neutralinos and involve
potentially large CP-violating phases originating from the mixing.
For the charginos, the CP-violating Lagrangian in the symmetric phase is
\begin{equation} 
{\cal L} = -g H^0_1 \bar{\tilde{H}} P_L \tilde{W} -
g H^0_2 \bar{\tilde{W}} P_L \tilde{H} + h.c.
\end{equation}
The CP-violating phases $\varphi_{\mu}$ and $\varphi_2$ are introduced
by  switching to the basis of mass eigenstates in the broken phase 
(see Appendix).  
Following the steps in \cite{riotto} to generate the CP-violating source term 
in the Boltzmann equation, we obtain for any point $X$ inside the bubble wall
\begin{equation}
{\cal S}_C  = 2g^2 {\rm Im}(M_2 \mu) v^2 (X) \dot{\beta} (X) 
{\cal I}_{\tilde{W}} \\
 = \gamma_{\tilde{W}} \sin(\varphi_{\mu} + \varphi_{2}),
\end{equation}
where  $v^2=v_1^2+v_2^2$ and 
$\dot{\beta} (X)=d\beta (X)/dt$ characterizes the temporal variation
of the Higgs profile $\tan\beta(X)=v_2(X)/v_1(X)$ as the wall passes through 
point $X$. ${\cal I}_{\tilde{W}}$ is a temperature dependent phase-space 
integral (the explicit form can be found in \cite{riotto}) and includes
information about thermal behavior of the winos and Higgsinos in the
high temperature plasma.  

The neutralino CP-violating interactions are
\begin{equation}
{\cal L}= -{1 \over 2} [ H^0_1 \bar{\tilde{H_1}} P_L (g_2 
\tilde{W_3} -g_1 \tilde{B}) + H^0_2 (g_2 \bar{\tilde{W_3}} -
g_1 \bar{\tilde{B}}) P_L \tilde{H_2} ] + h.c.
\end{equation}

Following the same steps as in the chargino case, we find the
source term
\begin{eqnarray}
{\cal S}_N &=& g_2^2 {\rm Im}(M_2 \mu) v^2 (X) \dot{\beta} (X) {\cal I}_{\tilde{W}}
+ g_1^2 {\rm Im}(M_1 \mu) v^2 (X) \dot{\beta} (X) {\cal I}_{\tilde{B}} 
\nonumber  \\
&=&\gamma_{\tilde{W}} \sin(\varphi_{\mu} + \varphi_2)+\gamma_{\tilde{B}} 
\sin(\varphi_{\mu} + \varphi_1). 
\end{eqnarray}

We can combine the source terms for the charginos and neutralinos to obtain
the total source term for Higgsinos
\begin{equation}
{\cal S}_{\tilde{H}} = 3 \gamma_{\tilde{W}} \sin (\varphi_2 + \varphi_{\mu}) +
\gamma_{\tilde{B}} \sin (\varphi_1 + \varphi_{\mu})
\end{equation}
where 
\begin{eqnarray}
\gamma_{\tilde{W}}&=& |\mu||M_2|g_2^2 v^2 (X) \dot{\beta} (X) 
{\cal I}_{\tilde{W}}, \\
\gamma_{\tilde{B}} &=& |\mu||M_1|g_1^2 v^2 (X) \dot{\beta} (X) {\cal I}_{\tilde{B}}.
\end{eqnarray}.

Here we emphasize that the source dependends on the full physical
(reparametrization invariant) 
combinations of the phases $\varphi_1 + \varphi_{\mu} $ and $\varphi_2 +
\varphi_{\mu}$ which factorize from the rest of the source term. In this
sense our considerations are independent of the particular details going
into the calculation of the Higgsino thermal production rate.   

The phase space integrals ${\cal I}_{\tilde{W}}$ and ${\cal
I}_{\tilde{B}}$ exhibit strong resonant behavior leading to a maximum
for $m_{\tilde{H}}\sim m_{\tilde{W}}$ ($m_{\tilde{H}}\sim
m_{\tilde{B}}$) \cite{riotto}. In terms of the soft breaking parameters
the enhancement occurs when  the gaugino masses $M_1$ or $M_2$ are
close in value to the Higgsino mass parameter $\mu$. For similar reasons
it is easy to understand why the right handed stop contribution to the
CP-violating source is always subdominant. The corresponding phase-space 
integral  ${\cal I}_{\tilde{t}_R}$ is typically far away from its
maximum as $m_{\tilde{t}_L}\gg m_{\tilde{t}_R}$ in our framework and the
stop source can therefore be neglected. 
 
The baryon asymmetry can be directly related to the Higgsino source by
solving a set of coupled diffusion equations for the Higgs and Higgsino,
top and stop, and the first two generation quark and squark densities
\cite{huetnel}.  
We assume that strong sphaleron transitions and interactions induced by the
top Yukawa coupling are very fast, allowing us to reduce the number of
relevant equations to the diffusion equations for Higgses and Higgsinos 
and baryon number.  They can be solved analytically, with the result
\begin{equation}
{n_B \over s} = - {81{\cal A} \bar{D} \Gamma_{ws} \over 82 v_w^2 s},
\end{equation}
where $\bar{D}$ is an effective diffusion constant, $\Gamma_{ws}$ is the
weak sphaleron transition rate, and $v_w$ is the velocity of the bubble wall.
The coefficient ${\cal A}$ is determined by boundary conditions to be
\begin{equation}
{\cal A} = {1 \over \bar{D} \lambda_+} \int_0^{\infty} du \ 
{\cal S}_H \  e^{-\lambda_+ u},
\end{equation}
where  
\begin{equation}
\lambda_+ = {v_w + \sqrt{v_w^2 + 4 \tilde{\Gamma} \bar{D}} \over
2 \bar{D}},
\end{equation}
and $\tilde{\Gamma}$ is an effective decay constant.  

 \section{Calculation of Washout}

Once the phase transition has occurred, weak sphaleron transitions tend to
erase any asymmetry that has been created.  In the broken phase, the weak
sphaleron rate is \cite{quiros}
\begin{equation}
\Gamma_{ws} \simeq 2.8 \times 10^5 T ({\alpha_W \over 4\pi})^4
\kappa ({E_{sp} \over B T})^7 e^{-E_{sp} \over T},
\end{equation}
where 
\begin{equation}
E_{sp} = {g v(T) B \over \alpha_W},
\end{equation}
and
\begin{equation}
B = B({\lambda \over g^2}) \simeq 1.87.
\end{equation}
$\kappa$ is a constant in the range $10^{-4} \leq \kappa \leq 10^{-1}$.  The
differential equation satisfied by $n_B$ in the broken phase is
\begin{equation}
{dn_B \over dt} = -C n_f \Gamma_{ws} n_B,
\end{equation}
where C is a number of $\cal O$(1) \cite{gmoore} and  we take
$C\simeq1$ absorbing the uncertainty into the uncertainty of $\kappa$
and $B$. Solving (14) we get
\begin{equation}
n_B (t) = n_B(t_c) \exp (-\int_{t_c}^t dt \  n_f \Gamma_{ws}).
\end{equation}

The traditional bound is found by assuming that all of the washout happens
at $T = T_c$, which amounts to saying that the integral in equation (3.5) is
equal to the value of the integrand at $T = T_c$.  However, we want to
evaluate the integral more carefully, and in the process derive a new bound.

Since we do not have an analytic formula for $v(T)$, we need to examine the
integrand to see if we can make any simplifying assumptions.  The sphaleron
rate is proportional to $(v(T)/T)^7 \exp(-36 v(T)/T)$, and we know that
$1 \lsim v(T)/T < \infty$. If 
we approximate that $v(T) = v(T_c)$ throughout this period, then the washout
rate is negligible when $T \simeq .75 T_c \simeq 75\, {\rm GeV}$.  
Since $v(T)$ changes by about a factor of 2.5 in the temperature range 
from (0\ - 100) GeV, treating $v(T)$ as a constant is a good
approximation; it is also conservative, 
because it will actually overestimate the washout slightly.

We can make a change of variables to
\begin{equation}
\zeta = {E_{sp} (T_c) \over T}
\end{equation}
by using the relation between time and temperature,
\begin{equation}
t \simeq (3 \times 10^{-2}) {M_{Pl} \over T^2}.
\end{equation}
The integral we now have to do is
\begin{eqnarray}
I &=& (3.4 \times 10^{-8}) {M_{Pl} \over E_{sp} (T_c)} \kappa \int_{\zeta_c}^{\infty}
\  \zeta^7 e^{-\zeta} d\zeta \nonumber \\
&=& (3.4 \times 10^{-8}) {M_{Pl} \over E_{sp} (T_c)} \kappa e^{-\zeta_c} \times
\sum_{n = 0}^7 {7! \over n!} \zeta_c^n
\end{eqnarray}
Because $\zeta_c = 4\pi B v(T_c) /gT_c \sim 36$, the $\zeta_c^7$ term will 
dominate all of the
other terms.  (Again, dropping the smaller terms is a conservative 
approximation.)  Then
\begin{eqnarray}
I &\simeq& (3.4 \times 10^{-8}) {M_{Pl} \over T_c} \kappa e^{-\zeta_c} 
\zeta_c^6 \nonumber \\
&\simeq& (4.1 \times 10^9) \kappa \zeta_c^6 e^{-\zeta_c}
\end{eqnarray}

Inserting this expression into equation (15), we now have an approximate 
expression for the observed $n_B/s$:
\begin{equation}
n_B (T \sim 0)/s = (n_B(T = T_c)/s) \exp (-(4.1 \times 10^9) \kappa \zeta_c^6 
e^{-\zeta_c}) \gsim 4 \times 10^{-11},
\end{equation}
from which we can obtain
\begin{equation}
\zeta_c - 6 \log \zeta_c - \log \kappa - 9 \log 10 - \log 4.1  + 
\log(\log {n_B/s (T_c) \over 4 \times 10^{-11}}) \gsim 0.
\end{equation}
By choosing a certain value for $\kappa$ and $n_B/s (T_c)$, this equation
can be solved numerically.  We will do this in the next section after
we have presented our numerical result for $n_B/s (T_c)$.  Once this is
done, we can use the relation
\begin{equation}
\zeta_c \simeq 36 {v(T_c) \over T_c}
\end{equation}
to find the lower bound on $v(T_c)/T_c$.

This lower bound on $v(T_c)/T_c$ can be translated into an upper bound on the
light Higgs mass.  The one-loop finite-temperature effective potential has
a form \cite{beqz,cqw1}
\begin{equation}
V(\varphi) = -{1\over 2}m^2 (T) \varphi^2 - T[E_{SM} \varphi^3 + F_{MSSM} (\varphi,T)]
+ {1\over 8} \lambda (T) \varphi^4.
\end{equation}
For the MSSM with heavy decoupled stops, the potential becomes SM-like 
and one has
\cite{beqz}
\begin{equation}
{v(T_c) \over T_c} \simeq {2 E_{SM} \over \lambda},
\end{equation}
where
\begin{equation}
E_{SM} = {2 M_W^3 + M_Z^3 \over 4\pi v^3}.
\end{equation}
Using $m_h^2 = 2 \lambda v^2$, equation  becomes
\begin{equation}
m_h^2 \lsim {4E_{SM} v^2 \over v(T_c)/T_c}.
\end{equation}
The other extreme is a light right-handed stop whose temperature
dependent self energy, responsible for screening of the stop interactions
in the plasma, is balanced by a negative soft squared mass term  
$-m_U^2 \sim
\Pi_{\tilde{t}_R}=(\frac{4}{9}g_3^2+\frac{4}{9} g^{\prime 2}) T^2
+\frac{1}{6}h_t^2 [1+\sin^2\beta (1-{\tilde{A}^2_t \over m^2_Q})] T^2$.
In such a case $F_{MSSM}$ is \cite{cqw1}
\begin{eqnarray}
F_{MSSM} (\varphi, T) &=& \varphi^3 E_{MSSM} \\
&=& \varphi^3 {m_t^3 (1-{\tilde{A}^2_t \over m^2_Q})^{3\over 2} \over 
2 \pi v^3},
\end{eqnarray}
where $\tilde{A}_t=A_t-\mu/\tan\beta$ is the stop left-right mixing 
parameter. Then
\begin{eqnarray} 
{v(T_c)\over T_c} &\simeq& {2 (E_{SM} + E_{MSSM}) \over \lambda},\\
m_h^2 &\lsim& {4 (E_{SM} + E_{MSSM}) v^2 \over v(T_c)/T_c}.
\end{eqnarray}
When screening is present ($m_U^2 +\Pi_{\tilde{t}_R} > 0$) the
temperature dependent Higgs potential can be analyzed numerically. 

\section{Numerical Results}

For the calculation of the baryon asymmetry, we have to adopt a model
for the electroweak phase transition and evaluate  $v^2 (X)$
and $\beta (X)$.  These functions have been calculated numerically using
the two-loop finite-temperature effective potential in \cite{moreno}.  However,
we have instead chosen to use the model of \cite{cqrvw} which lends itself
more easily to our computation:
\begin{eqnarray}
v(X) &=& v(L_w) [1 - \cos ({X\pi \over L_w})][\theta(X) -\theta(X-L_w)]
+ v(L_w)\theta(X-L_w), \\
\beta (X) &=& \Delta \beta [1 - cos({X\pi \over L_w})][\theta(X) -\theta(X-L_w)]
+ \beta(X = 0) + \Delta \beta \theta(X - L_w).
\end{eqnarray}
We have tested that the answer only weakly
depends on the model used for the phase transition, so the results should be
of at least the right order of magnitude.  We chose the value $\Delta \beta$
= .001 which is suggested by the results of \cite{moreno}, and a value
$v(L_w) \sim 100 \ {\rm GeV}$.  

The width of the wall is chosen to be  $L_w \sim
25/T$ \cite{cqrvw} and the typical velocity of the bubble wall is
expected to be $v_w\sim 0.1$.  
For the averaged diffusion coefficient we use $\bar{D} =$ 0.8 Ge${\rm
V^{-1}}$ and $\tilde{\Gamma} =$ 1.7 GeV as in \cite{riotto}. All of
these parameters enter the calculation of the baryon asymmetry, however,
the scaling dependence of $n_B$ on these quantities is straightforward and does
not interfere with CP-violation effects. 

\begin{figure}[h!]
\centering
\epsfxsize=4.75in
\hspace*{0in}
\epsffile{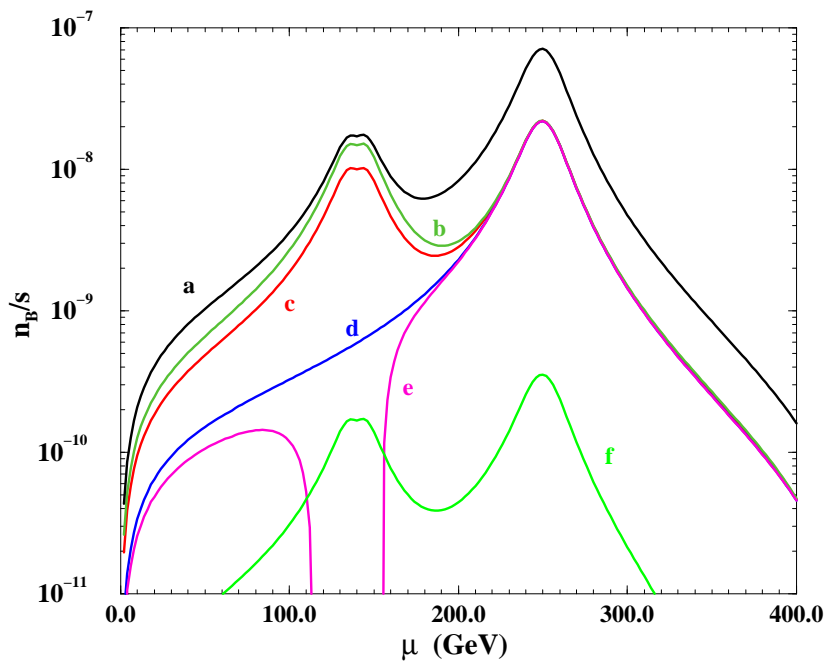}
\bigskip
\caption{Plot of the baryon asymetry produced at the electroweak phase 
transition for $|M_1| = 140\, {\rm GeV}$, $|M_2| =250 \,{\rm GeV}$, 
$\tan\beta\sim 3$ and $\varphi_{2}=0$. The five cases correspond to
some typical cases, with
{\bf a} $\varphi_{\mu}=\frac{\pi}{2}$, $\varphi_{1}=\frac{\pi}{10}$,
{\bf b} $\varphi_{\mu}=\frac{\pi}{10}$, $\varphi_{1}=\frac{\pi}{4}$,
{\bf c} $\varphi_{\mu}=\frac{\pi}{10}$, $\varphi_{1}=\frac{\pi}{10}$,
{\bf d} $\varphi_{\mu}=\frac{\pi}{10}$, $\varphi_{1}=-\frac{\pi}{10}$,
{\bf e} $\varphi_{\mu}=\frac{\pi}{10}$, $\varphi_{1}=-\frac{\pi}{8}$,
{\bf e} $\varphi_{\mu}=5\times 10^{-3}$, $\varphi_{1}=5\times 10^{-3}$.
}
\label{figone}
\end{figure}

Figure 1 shows the baryon asymmetry $n_B/s$ generated as a function of
$|\mu|$ for several combinations of $\varphi_{\mu}$ and $\varphi_{1}$
and $\varphi_{2}=0$, $|M_1| = 140\, {\rm GeV}$, $|M_2| =
250 \,{\rm GeV}$, $\tan\beta\sim 3$. The asymmetry     
$n_B/s$ can be as large as $10^{-7}$ for $|\mu| \sim |M_{2}|$.  

Let us note in passing that the signs of $\varphi_{1}$ and most
importantly of $\varphi_{\mu}$ (as it appears in both chargino and
neutralino matrices) determines the sign of $n_B$. This fact is also
demonstrated in Fig. 1 since $n_B$ is positive for $\sin\varphi_{\mu}>0$
with the exception of case {\bf e} and $\mu \sim M_1$ where $n_B$ turns 
negative being dominated by the neutralino contribution to the Higgsino 
source. Experimental determination of the magnitude {\it and} sign of
the soft phases is therefore essential if the feasibility of the
supersymmetric electroweak baryogenesis is to be verified. Alternatively,
the observed sign of $n_B$ could be used to determine the sign of
$\varphi_{\mu}$ until it can be measured other ways.

\begin{figure}[h!]
\centering
\epsfxsize=4.75in
\hspace*{0in}
\epsffile{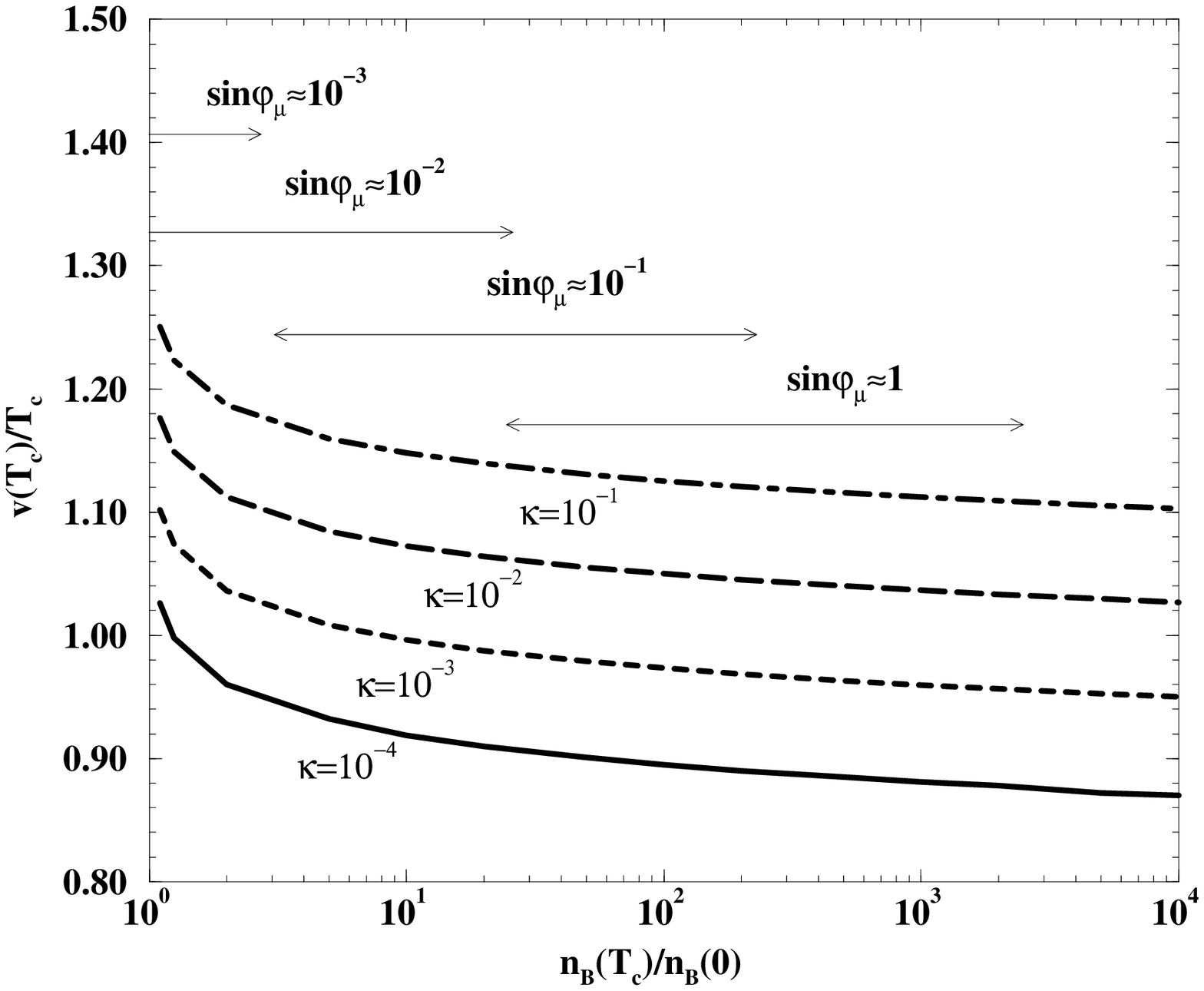}
\bigskip
\caption{Plot of required $v(T_c)/T_c$ ratio as a function of the actual
amount of baryon asymmetry produced at the electroweak phase transition.
The intervals for different values of $\sin\varphi_{\mu}$ illustrate the enhancement
resulting from quantum memory effects with the left (right) arrow
corresponding to no (full) quantum enhancement.  
}
\label{figtwo}
\end{figure}

In order to calculate the lower bound on $\zeta_c$, we have to specify
the the sphaleron parameters $B$ and $\kappa$. The non-perturbative
scaling factor $B$ comes from numerical minimization of the sphaleron
energy and was evaluated for the MSSM in \cite{moroakq}. The usual range
depending on the coupling strengths is $1.5<B(\frac{\lambda}{g^2})<2.7$
with a typical median of 1.87. The value of $\kappa$ is obtained as a
functional determinant associated with fluctuations about the sphaleron
and was estimated to be $\kappa \sim 0.1$ \cite{dine}. However, when the
Higgs propagator uncertainty is absorbed the allowed range for kappa is
$10^{-4}<\kappa<10^{-1}$.

The  required value of $v(T_c)/T_c$ depends on the amount of
dilution of the baryon number produced during the electroweak phase
transition that can be allowed in order to explain the observed value of 
$n_B/s$. Inclusion of both large CP-violating soft phases and 
quantum memory effects leads to a substantial enhancement in the
value of the produced asymmetry $n_B/s(T_c)$. Consequently, the
constraints on the strength of the first order phase transition can be 
softened as a function of $n_B(T_c)/n_B(0)$. 
This decrease of $v(T_c)/T_c$ is ilustrated in Fig. 2 for several
values of $\kappa$.  The enhancement in $n_B(T_c)$ coming from quantum
effects (about a factor of $10^2$ \cite{riotto}) for typical order of
magnitude values of $\sin\varphi_{\mu}$ is demonstrated by the
intervals with the position of the left (right) arrow corresponding to no
(full) quantum enhancemnent respectively. The minimal value of 
$\sin\varphi_{\mu}$ required in order to generate  $n_B/s(T_c)\sim
n_B/s(0)$ (negligible washout) without including the quantum effects in
the sources is about $5\times 10^{-2}$ which is in agreement with the
values obtained in Ref. \cite{cqw1}.
 
Since we are
looking for an absolute lower bound on $v(T_c)/T_c$, we will take 
$n_B/s (T_c)$ to have its maximum value, $10^{-7}$.  
For $\kappa = 10^{-1}$, the numerical solution
gives
\begin{equation}
\zeta_c \gsim 39.6,
\end{equation}
while the solution for $\kappa = 10^{-4}$ yields
\begin{equation}
\zeta_c \gsim 31.3.
\end{equation}
This translates into a bound on $v(T_c)/T_c$ of
\begin{eqnarray}
{v(T_c) \over T_c} &\gsim& 1.1,\ \  \kappa = 10^{-1}, \\
{v(T_c) \over T_c} &\gsim& .87, \ \ \kappa = 10^{-4}.
\end{eqnarray}
Using the bound for $\kappa = 10^{-4}$, we can find the upper bound on the
light Higgs mass.  For heavy stops, the upper bound is
\begin{equation}
m_h \lsim 51.7\, {\rm GeV}.
\end{equation}
For light stops with no thermal screening and negligible 
mixing ($\tilde{A}^2_t << m_Q^2$) the upper bound is
\begin{equation}
m_h \lsim 138.1\, {\rm GeV}.
\end{equation}

In Figure 3 we show a plot of the upper Higgs mass limit as a function of
the right handed stop mass calculated in our framework. The stop mass
range corresponds to variation of the right handed stop soft mass
parameter $m_U^2$ from $-\Pi_{\tilde{t}_R}$
(no thermal screening) to $\Pi_{\tilde{t}_R}$ (strong thermal
screening). The wide band corresponds to variation of $B$
and $\kappa$ within their full range and the narrow central band shows the
set of curves resulting from taking $B=1.87$ and varying $\kappa$ in the
full range. 

\begin{figure}[h!]
\centering
\epsfxsize=4.75in
\hspace*{0in}
\epsffile{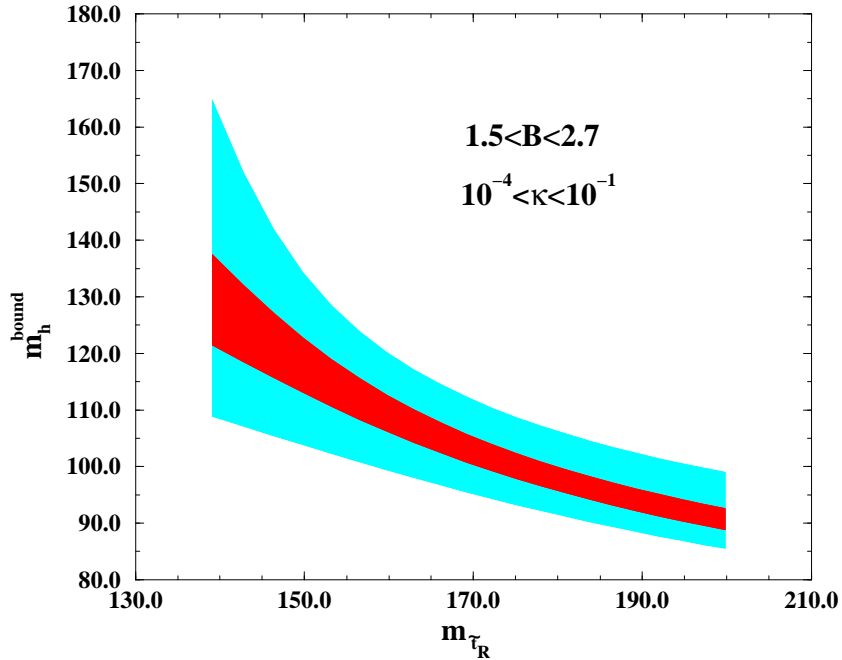}
\bigskip
\caption{Dependence of the upper Higgs mass bound on the light right
handed stop mass. The width of the bands results from uncertainties in
the  weak sphaleron parameters B and $\kappa$ which are varied in the
full range. The central band corresponds to B=1.87 while $\kappa$ is varied in
the full range.}
\label{figthree}
\end{figure}

It is important to stress that the role of large CP-violating phases is
crucial in this context. If the baryon asymmetry is overproduced during
the electroweak phase transition due to $\cal O$(1) values of
$\varphi_{\mu}$ and/or $\varphi_{1}$ the washout conditions (35) and
(36) are less stringent then if the phases are constrained to be
$10^{-2}$. As a result, the upper bound on the Higgs mass can be as much
as 15 GeV higher compared to the situation where $n_B/s(T_c)\sim n_B/s(0)$ and
no washout is allowed.  
 
It is obvious from our results that inclusion of large CP-violating
phases in the calculation of the baryon asymmetry relaxes the stringent
constraints on the strength of the first order transition and
consequently the light Higgs masses can be pushed towards larger values. 
Even for stop masses $m_{\tilde{t}_R}\gsim 170$ GeV resulting from  
positive values of the soft breaking parameter  $m_{U}^2$ the upper
bound on the Higgs mass can be as high as 115 GeV.  

Our considerations are based on the one loop temperature dependent 
effective Higgs potential evaluation. Two loop corrections are known to
significantly enhance the strength of the first order phase transition 
\cite{espydc}  and further relax the upper bound on the Higgs mass. In this
respect our results represent a conservative estimate of the upper Higgs
mass limit and it is likely to be moved upward when two loop
corrections are included.

\section{Summary and Conclusions} 

The upper bound on the Higgs mass which still allows electroweak baryogenesis
is an important issue which is becoming relevant as the Higgs mass experimental
lower limits are approaching 100 GeV. The baryon asymmetry produced
during the electroweak phase transition can succesfully explain the observed
value provided there are additional degrees of freedom contributing to
the finite temperature Higgs potential. Supersymmetric models with a light
right handed stop are a very good candidate for a theory that can provide
these degrees of freedom. Also, the CP-violating phases appearing in the
soft breaking terms of the Lagrangian can supplement (or entirely
replace) the effects of a CP-violating phase in the CKM matrix. 

Previously it has been thought that based on the electron and neutron
EDM experimental limits the supersymmetric CP-violating phases have to be small
and consequently there is no room for washout of the produced baryon
asymmetry. This translates into stringent constraints on the Higgs and
right handed stop masses, often leading to problems with color breaking
minima and potential stability.  
   
We have shown that once there is a possibility of cancellations among
individual contributions to the EDMs and the CP-violating phases are
allowed to be
$\cal O$(1), the produced baryon asymmetry exceeds the observed value
and the baryon density can be allowed to be diluted by a factor of
$10^{-3}$ or more. In such a case the upper Higgs mass bound is increased
and depending on the right handed squark mass it can go beyond 115-120 GeV
while at the same time the scalar potential is stable.

This scenario opens new possibile implications for supersymmetric
phenomenology. As pointed out in \cite{bk} it is important to
independently measure the CP-violating phases to correctly interpret
experimental observables. If the phases are small the window for
supersymmetric electroweak baryogenesis will get increasingly
smaller as the LEP and Tevatron experiments will be pushing up the lower
Higgs mass limit. It is natural to expect that in the case of small
phases the experimentally determined Higgs mass should not be too far
above 100 GeV if electroweak baryogenesis is expected to work. On the
other hand, if the Higgs is not discovered at LEP nor at the Tevatron, one
can expect that the CP-violating  phases of the MSSM are of $\cal O$(1)
if baryogenesis occurs at the electroweak phase transition,
and they should be measurable if the superpartners are discovered.
Of course, finding a light Higgs boson at LEP or Fermilab is consistent
with having large supersymmetric soft phases. 

\section{Acknowledgments}

We thank Jim Cline for valuable discussions, A. Riotto for 
helpful clarifications, and L. Everett for suggestions and comments on 
the manuscript. 
We also thank M. Quiros for correspondence.

\section{Appendix}
\def\theequation{A\arabic{equation}}
\setcounter{equation}{0}

Above the electroweak scale, the chargino mass matrix is
\begin{eqnarray}
{\cal M}_C \simeq \left(\begin{array}{cc}
 {|M_2| e^{i{\varphi}_2}} &
 0 \\
 0 &
 {|\mu | e^{i{\varphi}_{\mu}}}
 \end{array}\right).
\end{eqnarray}
This matrix is made real and diagonal by two complex matrices,
\begin{equation}
{\cal M}^{diag}_C = U^* {\cal M}_C V^{-1},
\end{equation}
where we can take
\begin{eqnarray}
U = \left(\begin{array}{cc} 1 & 0 \\ 0 & 1 \end{array}\right),
V = \left(\begin{array}{cc} 
 e^{i{\varphi}_2} & 0 \\
 0 & e^{i{\varphi}_{\mu}}
 \end{array}\right).
\end{eqnarray}
If we switch to a basis of two-component Weyl spinors, 
\begin{equation}
P_L \tilde{W} = P_L V^*_{j1} \tilde{\psi_j}, \ \ \
P_R \tilde{W} = P_R U_{j1} \tilde{\psi_j},
\end{equation}
\begin{equation}
P_L \tilde{H} = P_L V^*_{j2} \tilde{\psi_j}, \ \ \
P_R \tilde{H} = P_R U_{j2} \tilde{\psi_j},
\end{equation}
where $\tilde{\psi_j}$ are two-component spinors.  In terms of 
$\tilde{\psi_j}$, the interaction becomes
\begin{equation}
{\cal L} = -gH^0_1 e^{-i\varphi_2} \bar{\psi_2} P_L \psi_1 -
gH^0_2 e^{-i\varphi_{\mu}} \bar{\psi_1} P_L \psi_2 + h.c.
\end{equation}

Above the electroweak scale, the neutralino mass matrix is
\begin{eqnarray}
{\cal M}_N \simeq \left(\begin{array}{cccc}
 |M_1| e^{i\varphi_1} & 0 & 0 & 0 \\
 0 & |M_2| e^{i\varphi_2} & 0 & 0 \\
 0 & 0 & 0 & -|\mu| e^{i\varphi_{\mu}} \\
 0 & 0 & -|\mu| e^{i\varphi_{\mu}} & 0 \\
 \end{array}\right) .
\end{eqnarray}
It is made real and diagonal by the complex matrix ${N}$,
\begin{equation}
{\cal M}^{diag}_N = N^{\dag} {\cal M}_N N,
\end{equation}
where we can take
\begin{eqnarray}
N = \left(\begin{array}{cccc}
 e^{i\varphi_1 \over 2} & 0 & 0 & 0 \\
 0 & e^{i\varphi_2 \over 2} & 0 & 0 \\
 0 & 0 & {i \over \sqrt{2}} e^{i\varphi_{\mu} \over 2} & 
 {i \over \sqrt{2}} e^{i\varphi_{\mu} \over 2} \\
 0 & 0 & {1 \over \sqrt{2}} e^{i\varphi_{\mu} \over 2} &
 {-1 \over \sqrt{2}} e^{i\varphi_{\mu} \over 2}\\
 \end{array}\right).
\end{eqnarray}
Switching to the two-component spinor basis, we have
\begin{equation}
P_L \tilde{H_1} = P_L {N}^*_{j3} \tilde{\psi^0_j}, \ \ \
P_R \tilde{H_1} = P_R {N}_{j3} \tilde{\psi^0_j},
\end{equation}
\begin{equation}
P_L \tilde{H_2} = P_L {N}^*_{j4} \tilde{\psi^0_j}, \ \ \
P_R \tilde{H_2} = P_R {N}_{j4} \tilde{\psi^0_j},
\end{equation}
\begin{equation}
P_L \tilde{W_3} = P_L {N}^*_{j2} \tilde{\psi^0_j},\ \ \ 
P_R \tilde{W_3} = P_R {N}_{j2} \tilde{\psi^0_j},
\end{equation}
\begin{equation}
P_L \tilde{B} = P_L {N}^*_{j1} \tilde{\psi^0_j},\ \ \
P_R \tilde{B} = P_R {N}_{j1} \tilde{\psi^0_j},
\end{equation}
resulting in the interaction
\begin{eqnarray}
{\cal L} = &-& {g^2_2 \over 2\sqrt{2}} 
e^{-i(\varphi_2 + \varphi_{\mu})\over 2} 
[-iH^0_1 \bar{\tilde{\psi^0_3}} P_L \tilde{\psi^0_2} +
H^0_1 \bar{\tilde{\psi^0_4}} P_L \tilde{\psi^0_2} -
iH^0_2 \bar{\tilde{\psi^0_2}} P_L \tilde{\psi^0_3} -
H^0_2 \bar{\tilde{\psi^0_2}} P_L \tilde{\psi^0_4}]+ h.c.   \nonumber\\
&+& {g^2_1 \over 2\sqrt{2}} e^{-i(\varphi_1 + \varphi_{\mu}) \over 2}
[-iH^0_1 \bar{\tilde{\psi^0_3}} P_L \tilde{\psi^0_1} +
H^0_1 \bar{\tilde{\psi^0_4}} P_L \tilde{\psi^0_1} -
iH^0_2 \bar{\tilde{\psi^0_1}} P_L \tilde{\psi^0_3} -
H^0_2 \bar{\tilde{\psi^0_1}} P_L \tilde{\psi^0_4}] + h.c.
\end{eqnarray}   

\def\B#1#2#3{\/ {\bf B#1} (19#2) #3}
\def\NPB#1#2#3{{\it Nucl.\ Phys.}\/ {\bf B#1} (19#2) #3}
\def\PLB#1#2#3{{\it Phys.\ Lett.}\/ {\bf B#1} (19#2) #3}
\def\PRD#1#2#3{{\it Phys.\ Rev.}\/ {\bf D#1} (19#2) #3}
\def\PRL#1#2#3{{\it Phys.\ Rev.\ Lett.}\/ {\bf #1} (19#2) #3}
\def\PRT#1#2#3{{\it Phys.\ Rep.}\/ {\bf#1} (19#2) #3}
\def\MODA#1#2#3{{\it Mod.\ Phys.\ Lett.}\/ {\bf A#1} (19#2) #3}
\def\IJMP#1#2#3{{\it Int.\ J.\ Mod.\ Phys.}\/ {\bf A#1} (19#2) #3}
\def\nuvc#1#2#3{{\it Nuovo Cimento}\/ {\bf #1A} (#2) #3}
\def\RPP#1#2#3{{\it Rept.\ Prog.\ Phys.}\/ {\bf #1} (19#2) #3}
\def\etal{{\it et al\/}}
\bibliographystyle{prsty}

\newpage

\end{document}